\documentclass[useAMS,usenatbib,letterpaper]{mn2e}
\newcommand{\kms}{{\rm {km\, s^{-1}}}}
\newcommand{\cm}{{\rm {cm}}}
\newcommand{\yr}{{\rm {yr}}}

\newcommand{\msun}{M_{\odot }}
\newcommand{\msigma}{$M_{\bullet}$--$\sigma_*$\ }
\begin{document}
\title[Possible evidence for the ejection of a supermasive black hole]
{Possible evidence for the ejection of a supermassive black hole
from an  ongoing merger of galaxies}
\author[M. G. Haehnelt, M. B. Davies and M.J. Rees]{Martin G. Haehnelt$^{1}$\thanks{E-mail:
haehnelt@ast.cam.ac.uk (MGH); mbd@astro.lu.se (MBD); mjr@ast.cam.ac.uk (MJR)}, Melvyn B.
Davies$^{2}$ and Martin J. Rees$^{1}$\\
$^{1}$Institute of Astronomy, Madingley Road, Cambridge, CB3 OHA\\
$^{2}$Lund Observatory, Box 43, SE--221 00, Lund, Sweden}
\maketitle


\label{firstpage}

\begin{abstract}
Attempts of Magain et al (2005) to detect the host galaxy of the
bright QSO  HE0450--2958 have not been successful. We suggest that 
the supermassive black hole powering the QSO was ejected from 
the observed ULIRG at the same redshift and at  1.5  arcsec distance. 
Ejection could have either be  caused  by recoil due to gravitational 
wave emission from a coalescing binary 
of supermassive black holes or the gravitational slingshot of 
three or more supermassive black holes in the ongoing merger 
of galaxies which triggered the starburst activity in the ULIRG. 
We discuss implications for the possible  hierarchical build-up 
of supermassive black holes from intermediate and/or stellar 
mass black holes,  and  for the detection of coalescing supermassive 
binary black holes  by LISA.    
\end{abstract}


\section{Introduction}

In their recent paper, Magain et al (2005) describe observations
of the bright quasar HE0450--2958 which suggest that its host galaxy
is at least six times fainter than expected for the typical relation
of black hole mass to bulge luminosity.  Such efficient black hole
formation,  with  a large black hole mass surrounded by  a 
relatively small mass of stars,  would be somewhat surprising. In this {\sl letter}
we suggest an alternative explanation for their observation: namely 
that the supermassive black hole powering the quasar  has been ejected
from the centre of an ongoing merger of galaxies during the 
violent dynamical interaction of two or  more black holes. A plausible 
location for such an event is the  ultra-luminous
infra-red galaxy  (ULIRG) also described by Magain et al. 
which lies at a distance of $\sim$ 1.5 arcsec. 

ULIRGs are known  to be powered by bursts in star
formation activity which are often triggered by the merger of two 
(or more) gas-rich galaxies.  Each of the merging galaxies is expected 
to have contained at least one supermassive black hole with a mass
that scales roughly with the mass of the bulge of the merging galaxies
({\it e.g.} Kormendy \& Richstone 1995). The 
black holes in the merging galaxies will sink quickly to the centre of 
the merger product and will form a hard binary (Begelman, Blandford \&
Rees 1980 , Miloslavjevic \& Merritt 2001). The further evolution is somewhat
uncertain. Hardening by gravitational interaction with stars passing
close to the binary will be  much slower than  the typical duration of 
the star formation bursts  in ULIRGs. However, if the binary were
submerged in a gaseous disc which formed from gas funnelled to the centre during the ongoing
merger the separation of the  binary could shrink fast  to the
point where gravitational radiation leads to rapid coalescence
(Begelman et al. 1980). Of course the dynamical interaction could be more complex if one 
or both of the two merging galaxies contained binary or multiple supermassive 
black holes.   There are two basic gravitational processes that can
eject one or more of the  black holes from the merging galaxies, 
the gravitational radiation recoil due to asymmetric  emission of 
gravitational radiation and the
gravitational slingshot during the violent dynamical interaction of
three  or more black holes.  

This {\sl letter} is arranged as follows.
We will briefly assess the required
ejection velocity in section 2. We will then discuss the two ejection
mechanisms and their possible implications for the build-up
of supermassive black holes from intermediate mass and/or stellar mass
black holes  in sections 3 and 4. We discuss in section 5 how the ejected black
hole is likely to be supplied with material. 
In section 6 we summarize and discuss our results.

\section{The required kick velocity}

HE0450--2958 is at a redshift, $z=0.285$. Assuming standard cosmological
parameters, 1 arc second is equivalent to $\sim$ 4.3 kpc at that redshift.
In projection, the quasar is observed to be around 1.5 arc seconds away
from the centre of the companion galaxy. In order to see whether the 
black hole merger and ejection picture works, let us assume that 
the distance is in fact 10 kpc. 
Assuming an average velocity of  300 km/s, 
the black hole would take about 30 Myr to travel the required distance,
which is consistent with the picture that a starburst was triggered
by a merger involving the companion galaxy some 100 Myr ago (Canalizo
\& Stockton 2001). This velocity is consistent with the observed
radial velocity difference between the quasar and the companion
galaxy which is about 130 km/s (Magain et al 2005). Assuming a
lifetime of a the starburst phase in a ULIRG of $10^7$ to $10^8$ yr 
an average  velocity of 100-1000 km/s is required to travel the observed
distance.  Note, however that the black hole is likely to sit  deep 
in the potential well of the  dark matter halo hosting the ULIRG 
and therefore may have been decelerated.  We can only
speculate about the properties of the dark matter halo  hosting the 
ongoing merger of galaxies.  The rather large mass of the black hole 
powering HE0450--2958 suggests however that its mass and virial radius
are large. Ferrarese (2004)  has established an empirical relation 
between the circular velocity of the DM halos hosting galactic bulges
and the mass of the central supermassive black hole. If the black hole radiates at 50\% of
its Eddington luminosity as assumed by Magain et al. the  mass of the central
black hole is  $\sim 8 \times 10^8$ M$_\odot$. This would correspond
to a mass of the dark matter halo of $\sim 10^{13} \msun$ 
and a virial radius of $\sim$ 300 kpc.  These values are somewhat
larger than for typical ULIRGs (Tacconi et al. 2002).  
The escape velocity from such a dark matter halo  will be about  $\sim$ 1000 km/s.  
It appears thus indeed likely that the black hole is being decelerated and will not be able to
escape the DM halo unless the average velocity of the supermassive
black  hole in HE0450--2958 is at the upper limit of our estimate or we have
overestimated the  escape velocity of the DM halo. Obviously if the 
black hole is decelerated the current velocity will be lower than the average
velocity since ejection. Note further that unless the ejection 
velocity exceeds 50\%  of the escape velocity the black hole 
would be expected to fall back to the centre of the potential
well in $10^{8} \yr$ or less (Merritt et al 2004).

\section{Ejection by  gravitational radiation recoil} 

When two black holes of unequal mass merge, the merger product 
will receive a kick due to the asymmetric emission of gravitational radiation 
(Fitchett \& Detweiler 1984,   Redmount  \& Rees 1989). 
Full numerical simulations of this process are not yet
possible. Analytical calculations  rely on perturbation theory and are
somewhat uncertain for mass ratios of the merging black holes which
are not small.  The most recent calculations of this kind by 
Favata et al 2004 and Blanchet et al 2005 obtain values of 50 -- 300
km/s.   The exact value of the  kick velocity is a function of the black
holes masses and spins.  The major question mark
is posed by the required very short  merging time scale for the black holes. 
The interaction with a gaseous disc  could in principle lead to a 
fast merging of the black holes (Begelman et al. 1980, Armitage \&
Natarajan 2002, Escala et al. 2004).  If this is indeed what has
happened in  HE0450--2958 this is good
news for the planned space-based gravitational wave interferometer 
LISA\footnote{{\tt http://lisa.jpl.nasa.gov/}} which aims at
detecting  the merging of supermasive black hole
binaries albeit of somewhat smaller mass.   


The ejection of supermassive black hole has also very interesting 
implications for models for the joint hierarchical  build-up of 
supermassive black holes and galaxies (e.g. Kauffmann \& Haehnelt 2000).  
If HE0450--2958 was indeed ejected by recoil due to gravitational
radiation,  the recoil velocities  would have to lie towards  the upper
end of the range suggested by Favata et al. 
and Blanchet et al.  Binary mergers would then easily eject black
holes  from dwarf  galaxies and could  lead to the displacement 
of supermassive black holes to the outer parts of even the most
massive galaxies (Merritt et al. 2004).  
The hierarchical build-up of  supermassive black holes would then  
need to be highly fine tuned if it were to extend to stellar mass 
black holes. Volonteri et al (2002) have discussed such a model 
where the stellar mass black hole seeds are not much more numerous  
than the number of supermassive black holes in 
bright galaxies,  and binary mergers in small galaxies  are relatively
rare.  Note further that Haehnelt \& Kauffmann (2002, see also
Volonteri et al. 2003) have argued 
that the merging of supermassive black holes has to occur on
time scales of less than a Hubble time to avoid excessive scatter in 
the \msigma relation between black hole mass and stellar velocity
dispersion of galactic bulges (Gebhardt et al. 2000; 
Ferrarese \& Merritt 2000).  Otherwise there should be  
the occasional bright elliptical galaxy  with no (or a significantly 
under-weight) central black hole  which appears not to be the case 
(see  Merritt \&  Miloslavjevic 2004 for a review on 
supermassive binary black holes and evidence for their coalescence).

\section{Ejection by  gravitational slingshot} 
 
It is far from certain that the gas driven to
the centre of a ULIRG will actually lead to the fast merging of 
a supermassive binary.  
Most of the gas may actually be consumed at larger radii by the 
starburst or may  be accreted by the two black holes without efficiently hardening the
binary black hole. An alternative scenario for the ejection of a black
hole during a galaxy mergers is the  gravitational slingshot (Saslaw,
Valtonen \& Aarseth 1974, Hut \& Rees 1992). 
One or perhaps even both of the galaxies may already have 
contained a hung-up hard supermassive binary black hole. When the two galaxies
merge the  black holes will undergo a violent dynamical interaction. 

To be more specific consider the case of a binary plus a single black hole. The
most likely outcome is that the lightest of the black holes will be
flung out with a velocity comparable  to the circular velocity of the
hard binary. The two other black holes will  stay bound to each other
and receive a smaller kick to conserve the total momentum. 
Yu (2002) has used the stellar density profiles   of nearby galaxies
to estimate the most likely circular velocity of hung-up binaries
assuming that they contain supermassive binary black holes hardened by
scattering of stars. The results depend on details of the  stellar
orbits, assumptions about loss cone refilling  and the mass ratio of 
the binary. The values range  from 500-2000 km/s.  
This and the rather large mass of the black hole in HE0450--2958 make it then perhaps more likely
that  a lighter black hole of  $\sim 10^7-10^8\msun$  was flung out
at a speed comparable to the circular velocity of a hung-up binary and 
that the black hole  powering  HE0450--2958 is the binary that is 
ejected at somewhat lower speed in the opposite direction. 
If the lighter  black hole is much less massive 
than the binary the binary has to be very hard, 
if the binary is to be ejected with adequate speed. 
Scattering of stars  would  then be too slow to bring the 
third black hole close enough for the  gravitational slingshot 
during the lifetime of the ULIRG, but  interaction with a gas disc could drive 
the third black hole in on the required time  scale.   
Note that Yu (2002) will have underestimated 
the circular velocity at which  supermassive binary black holes 
in nearby galaxies should have got hung up if the hardening of the binaries  
has significantly affected the stellar distribution at the centre of these galaxies.  

If the supermassive black hole in HE0450--2958 was ejected by a
gravitational slingshot the possible implications for the hierarchical 
build-up of supermassive black holes are  less clear. 
Hardening by scattering of stars should be more efficient in small galaxies where 
loss cone refilling should occur on shorter time scales (Yu 2002). The presence of a hung-up
hard supermassive binary  is thus much less likely in small galaxies.  
The second ejection scenario  would therefore not necessarily be bad  
news for LISA which will be sensitive to black holes of smaller mass.

\section{Feeding HE0450--2958}

Magain et al claim that their deconvolution of the HST image of 
HE0450--2958 has revealed a blob of gas to one side of the point-like 
QSO emission opposite to the location of the ULIRG. The blob has a
diameter of  $\sim$ 2 kpc.  Its emission shows no sign of a  
stellar continuum, but is bright in $H_{\alpha}$, $H_{\beta}$ and $O_{III}$  
with line ratios consistent with no dust absorption.  
Magain et al. interpret this blob as an emission
nebula excited by the QSO emission.  
The obvious question is whether 
this blob is related to the feeding of the  SMBH in  HE0450--2958. 
The presence of such a gas cloud  at a distance of 10kpc of an  ULIRG 
is not implausible as the merging galaxies powering ULIRGs are gas rich and 
are surrounded by  substantial  amounts of tidal  debris.  Could 
this cloud actually feed the supermassive black hole    
in HE0450--2958?  If the black hole accretes  at the Bondi-Hoyle accretion rate 
it would have to accrete from material with a density $n_{\rm gas} 
\sim 15 \;(\eta/0.1)^{-1}\;(v_{\rm bh}/200 \kms)^{3} \cm^{-3}$  to produce the
observed luminosity,  where $\eta$ is the efficiency for turning rest
mass energy into radiation  and $v_{\rm bh}$ is the relative velocity 
between black hole and gas cloud. The supermassive black hole in 
HE0450--2958 must  have traveled  at an average velocity $\ga 100\kms$  
but as discussed in section 2 the current velocity could be smaller than the ejection
velocity.  If the supermassive black
hole in HE0450--2958 indeed sits in a potential well corresponding to 
a 1D velocity dispersion of 300 km/s as suggested by the \msigma
relation, then the mass contained in a 10kpc sphere would be $4\times 10^{11} \msun$.  
The gas distribution should be inhomogeneous. The 
probability for hitting material of the density for Bondi-Hoyle accretion at
the required rate would be $0.005 \; (v_{\rm bh}/200 \kms)^{-3} \; (M_{\rm
gas}/10^{10} \msun)  \;$ where $M_{\rm
gas}$ is the mass of gas at this density.  Even for a rather small
fraction  of the total mass in gas at the right density this is thus not
implausibly low given the number of observed 
ULIRGs. It would, however, imply that ejection of a
supermassive (binary) black hole is a common phenomenon in ULIRGs.  
The accretion from a blob of such density at a rate close to the 
Eddington rate should not be much different from the accretion
during the typical QSO phase of an accreting black hole at the centre of 
merging galaxies. Unfortunately, we would therefore  not expect a special spectral       
signature other than that obscuration is less likely and indeed on 
first sight the spectrum of HE0450-2958 appears to show no peculiarities.

An alternative  option is that 
the ejected black hole is fed by accretion of material
which remained bound to the black hole when it was ejected from the host
galaxy.  An ejected binary should carry with it  a substantial
amount of gas that is gravitationally bound to it. 
 Considering a sphere of material of total mass $10^9$ M$_\odot$, 
one notes that the surface escape speed exceeds 300 km/s for radii less
than $\sim$ 100 pc.  Any gas within this radius would
have been retained by the (merging) black hole(s) when  they received
a kick of 300 km/s. Interestingly, 100 pc is the upper limit Magain et
al quote for the half light radius for the emission directly
surrounding the QSO.  We therefore
conclude that the black hole should have been able to retain enough fuel to power
the quasar during the previous 30 Myr. The location of  HE0450--2958 
at the  edge of the observed gas blob would then  be most
plausibly explained as being due to the driving of a wind by the 
QSO activity.  In the case of ejection by a gravitational slingshot 
there is the intriguing but not very likely possibility of the detection of one
perhaps even two further QSOs on the other side of the ULIRG. The 
distance could be as large as an arcmin or more. These would almost certainly be 
significantly fainter and would have to have been ejected with 
sufficient fuel as  feeding a fast moving black hole 
at a large distance from the ULIRG would be very difficult otherwise.

\section{Summary and Discussion}

We have discussed here the possibility  that the supermassive black hole powering 
the bright quasar HE0450--2958 has been ejected during the ongoing merger of
galaxies responsible for the nearby ULIRG.  The projected distance
between the ULIRG 
and HE0450--2958 is consistent with an average speed 
of  300 km/s since ejection 30 Myr ago.  Feeding could be either by accretion
from the gas blob located next to HE0450--2958 on the opposite side
from the ULIRG or by accretion of material which was flung out 
from the centre of the merger together with the
black hole.  It is likely that the ejection velocity is not sufficient to
escape the surrounding dark matter halo and that the black hole has been decelerated. 
The two  plausible  ejection scenarios have interesting implications
for the possible hierarchical build-up of supermassive black holes
from intermediate and/or stellar mass black holes. If ejection occured 
by gravitational radiation recoil this would be the first
identification of the coalescence of a binary of supermassive black
holes. The coincidence of the coalescence of the 
black hole  binary and the  starburst activity in the ULIRG then 
suggests that both are causallly connected
and that the gas funneled to the centre during the merger of the
galaxies has led to the rapid hardening of a supermassive binary
that either existed in the merging galaxies before the merger or 
was formed during the merger.  Rapid merging of supermassive binary
black holes may then be widespread in merging galaxies. This  would be
excellent news for the planned space-based gravitational wave
interferometer  LISA. It may also argue for the formation of
supermassive black holes from rather massive seed black holes:
the alternative models, assuming the hierarchical build-up of supermassive black holes from 
intermediate and/or stellar mass black  holes would need considerable 
fine-tuning to avoid ejection from shallow potential wells at high redshift.  

Different conclusions are to be drawn if ejection was caused 
by a gravitational slingshot during the violent dynamical interaction 
of three or more black holes.  The occurence of such a slingshot would 
suggest that one or both of the galaxies contained a 
hung-up hard binary  and that accretion of gas may not be efficient in 
hardening supermassive binary black holes to the point of rapid
coalescence due to gravitational wave emission.  
There may then also be the small chance of the discovery of one or two
fainter QSOs on the other side  of the ULIRG. 
In the case of ejection by a gravitational slingshot  
LISA and models for the hierarchical build-up of supermassive black holes from 
intermediate and/or  stellar  mass black holes would have to rely on efficient hardening by
scattering of stars into the loss cone of supermassive binary 
black holes in shallow potential wells. 

We end with the caveat that further
scrutiny of HE0450-2958 may still reveal a very underluminous and
compact host galaxy, which would be interesting in its
own right.

\section*{Acknowledgments}

MBD is a Royal Swedish Academy Research Fellow supported by
a grant from the Knut and Alice Wallenberg Foundation.
We thank the Swedish Royal Academy of Sciences  for an invitation 
to the Crafoord symposium  2005 in Stockholm where this work 
was initiated by a presentation by David Elbaz. We further thank Mitch
Begelman for helpful comments on the manuscript.

\label{lastpage}

\end{document}